\documentclass[final]{cvpr}

\usepackage{times}
\usepackage{epsfig}
\usepackage{graphicx}
\usepackage{amsmath}
\usepackage{amssymb}
\usepackage{booktabs}
\usepackage{multirow}
\usepackage{caption}
\usepackage{textcomp}
\usepackage{siunitx}
\usepackage{color, colortbl}
\usepackage[english]{babel}
\usepackage{amsthm}

\usepackage{pifont}
\newcommand{\bfsection}[1]{\vspace*{0.1cm}\noindent\textbf{#1.}}

\usepackage[x11names]{xcolor}
\usepackage[pagebackref=false,breaklinks=true,letterpaper=true,colorlinks,bookmarks=false]{hyperref}
\hypersetup{
     colorlinks = true,
     linkcolor = red,
     anchorcolor = black,
     citecolor = SpringGreen4,
     filecolor = black,
     urlcolor = Firebrick1,
     }

\newcommand{\red}[1]{\textcolor{red}{#1}}

\def\cvprPaperID{xxxx} 
\def\confYear{CVPR 2021}

\begin{document}

\title{PyTorch Connectomics: A Scalable and Flexible Segmentation\\ Framework for EM Connectomics}

\author{%
Zudi Lin$^1$\thanks{Contact email: \texttt{linzudi@g.harvard.edu}} \quad Donglai Wei$^{1,2}$ \quad Jeff Lichtman$^1$ \quad Hanspeter Pfister$^1$\\[2mm]
$^1$Harvard University \quad $^2$Boston College
}

\maketitle
\begin{figure*}[t]
    \centering
    \includegraphics[width=\textwidth]{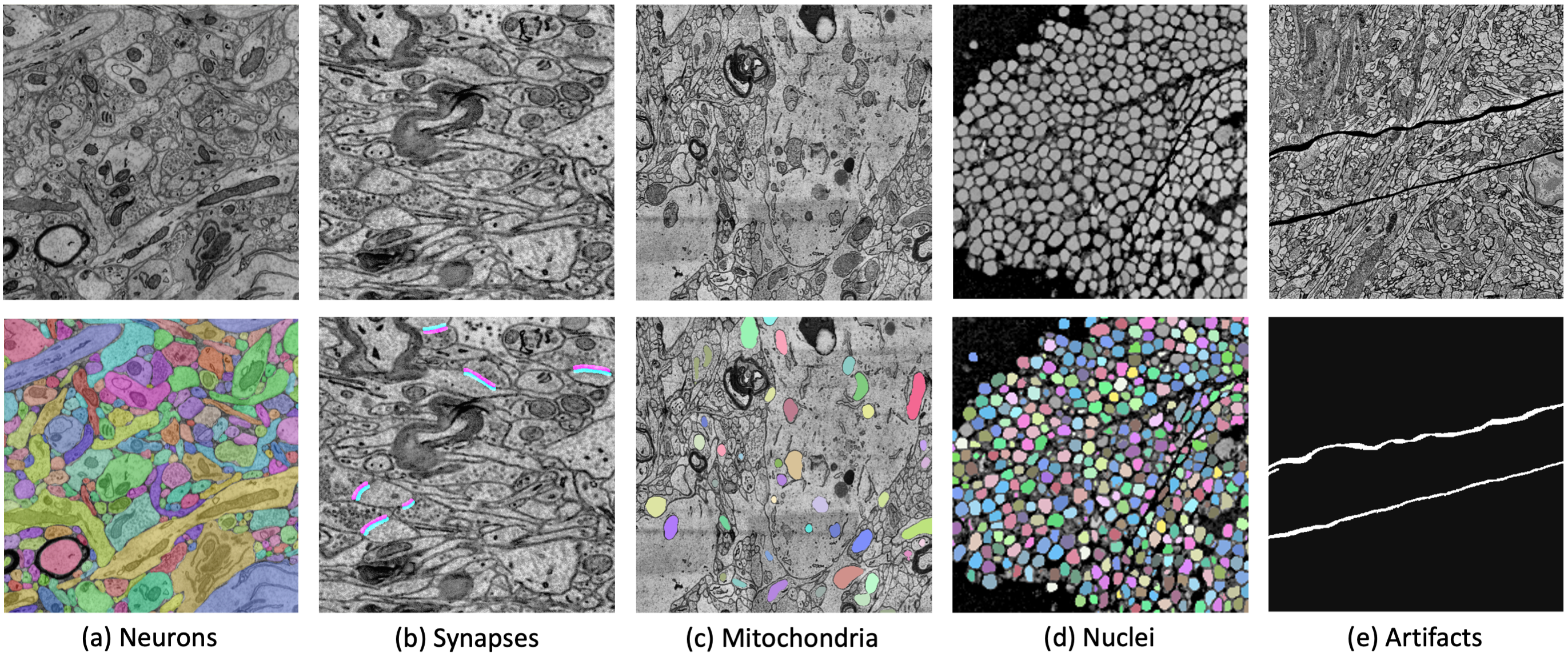}
    \caption{Illustration of supported tasks in {\em PyTorch Connectomics} (PyTC). Our package supports the semantic and instance segmentation of (a) neurons, (b) synapses, (c) mitochondria, (d) nuclei, and (e) tissue processing artifacts from microscopy images of animal brains. The versatile configuration options make the framework easily adaptable to various learning targets. 
    }\label{fig:teaser}
\end{figure*}
\begin{abstract}
We present PyTorch Connectomics (PyTC), an open-source deep-learning framework for the semantic and instance segmentation of volumetric microscopy images, built upon PyTorch.
We demonstrate the effectiveness of PyTC in the field of connectomics, which aims to segment and reconstruct neurons, synapses, and other organelles like mitochondria at nanometer resolution for understanding neuronal communication, metabolism, and development in animal brains. 
PyTC is a scalable and flexible toolbox that tackles datasets at different scales and supports multi-task and semi-supervised learning to better exploit expensive expert annotations and the vast amount of unlabeled data during training. Those functionalities can be easily realized in PyTC by changing the configuration options without coding and adapted to other 2D and 3D segmentation tasks for different tissues and imaging modalities. Quantitatively, our framework achieves the best performance in the CREMI challenge for synaptic cleft segmentation (outperforms existing best result by relatively 6.1$\%$) and competitive performance on mitochondria and neuronal nuclei segmentation. Code and tutorials are publicly available at \url{https://connectomics.readthedocs.io}.
\end{abstract}

\section{Introduction}

The brain is one of the most complicated and mysterious tissues, controlling the motion, perception, and emotion of animals (including us humans). However, the detailed brain structure of most animals remains largely unknown to date due to its sheer complexity. Our research focuses on connectomics, which aims to solve the comprehensive connectivity graphs in the animal brains to shed light on the underlying mechanism of intelligence and inspire better treatment for neurological diseases. Recently, with the development of modern electron microscopy (EM) techniques, neuroscientists have been collecting petabytes of volumetric images at {\em nanometer} resolution, enabling the analysis of detailed neuronal connectivities and subcellular structures like synapses and mitochondria~\cite{kasthuri2015saturated,xu2020connectome,shapson2021connectomic}. 

Segmenting cellular and subcellular structures from microscopy images is usually the first step before structural~\cite{talwar2020topological} and functional~\cite{shapson2021connectomic} analysis. In terms of computing, the segmentation tasks can be generally categorized into two classes, including {\em semantic} segmentation that assigns each pixel a class label and {\em instance} segmentation that additionally assigns each pixel an object index besides the class label. Nowadays, deep learning has become the {\em de facto} methodology for both semantic segmentation~\cite{chen2017deeplab} and instance segmentation~\cite{he2017mask,ronneberger2015u}, which has achieved state-of-the-art performance on most natural-scene and biomedical image datasets without using hand-designed features.

However, there are several challenges when applying those automatic approaches to the segmentation of large-scale connectomics datasets. First, existing open-source segmentation packages are usually designed for standard benchmark datasets, which means adapting those tools to new data with different scales and formats requires considerable time in changing hardware conditions and data pre-processing. Second, implementing specialized network architectures or different learning targets for new tasks is usually cumbersome in prevailing frameworks. Another layer of challenge is that most packages, although having good performance on segmentation tasks, lack detailed tutorials and documentation, which adds more learning burden for researchers without much experience in programming. 

To tackle the challenges, we present {\em PyTorch Connectomics} (PyTC), an open-source deep-learning framework for the semantic and instance segmentation of volumetric microscopy images. PyTC is built upon the state-of-the-art deep learning library PyTorch~\cite{NEURIPS2019_9015} and inherits its advantages in flexibility and simplicity for neural network implementation. Our framework is not designed for specific imaging modality, data scale, or model architecture but versatile for different kinds of datasets and segmentation tasks, which distinguish us from previous works like flood-filling networks (FFN)~\cite{januszewski2018high} for neuron segmentation, StarDist~\cite{weigert2020star} for nuclei segmentation, and CleftNet~\cite{liu2021cleftnet} that only focuses on synapses. Similar to nnU-Net~\cite{isensee2021nnu}, we identify a set of components like data loaders, image augmentations, and optimizers that require little adaptation among different datasets and tasks. However, nnU-Net only handles semantic segmentation, while our framework can tackle both semantic and instance segmentation. Furthermore, our framework is much more customizable than nnU-Net to tackle unique challenges in different datasets and tasks, which helps us achieve significantly better performance (\eg, synaptic cleft segmentation). Several supported tasks of PyTC are demonstrated in Figure~\ref{fig:teaser}.

Specifically, we design PyTC by identifying the primary needs of connectomics researchers in different components of segmentation. For data loading, we handle large datasets by automatically splitting them into manageable chunks, which is compatible with parallel and distributed processing and can easily be adapted to different hardware conditions. For modeling, researchers can customize 2D, 3D, or mixed architectures (for anisotropic images) and change the number of layers and kernels based on data complexity. The model can also learn multiple targets with multiple losses for each target, facilitating hybrid-representation learning to fully utilize annotations. Besides, users can easily configure data-loading, model-building, optimization, and visualization by just changing a configuration file without coding. The design also ensures the reproducibility of experiments and is friendly for the development and maintenance of the package. Finally, we provide a detailed description of the design and usage of PyTC in our documentation and tutorials to accelerate the learning curve.

To summarize, we present the {\em PyTorch Connectomics} framework for handling challenges in volumetric microscopy image segmentation by achieving scalability and flexibility in data processing, modeling, and learning. We have been actively improving the framework with feedback from biomedical researchers and help from open-source contributors. We are incorporating recent machine learning techniques like semi-supervised and self-supervised learning to effectively utilize vast unlabelled data, which can better assist experts in connectomics research.

\section{Related Work}
We here review popular image and volume segmentation packages for general microscopy and EM Connectomics.

\bfsection{Segmentation Packages for General Microscopy}
On the one hand, many segmentation packages are based on traditional computer vision and machine learning methods as surveyed in~\cite{moen2019deep,lucas2021open}. These packages have accelerated quantitative and statistical analyses for bacteria cells~\cite{stylianidou2016supersegger,paintdakhi2016oufti,ursell2017rapid}, mammalian cells~\cite{carpenter2006cellprofiler,sommer2011ilastik,belevich2016microscopy}, and general-purpose objects~\cite{schindelin2012fiji,allan2012omero}. Recently, some of these packages began to support deep learning methods~\cite{gomez2021deepimagej,mcquin2018cellprofiler}.
On the other hand, new packages strive to make state-of-the-art deep learning methods, \eg~U-Net~\cite{ronneberger2015u}, accessible to non-expert users in biology~\cite{haberl2018cdeep3m,van2016deep,wolny2020accurate}. %
Moreover, StarDist~\cite{weigert2020star} and Cellpose~\cite{stringer2021cellpose} built user-friendly packages upon their proposed novel cell segmentation representations for better performance. nnU-Net~\cite{isensee2021nnu} aims to learn to configure the best U-Net-like model for the given segmentation task. 
However, these segmentation packages, designed for general microscopy, do not work well for nanometer-scale electron microscopy images where the 3D cell structures are more complex and crowded. 
Our PyTC package focuses on the segmentation tasks for EM Connectomics.

\bfsection{Segmentation Packages for EM Connectomcis}
Most EM Connectomics segmentation packages are programmer-centric, only providing research code for their own proposed methods~\cite{berning2015segem,januszewski2018high,lee2017superhuman,sheridan2021local}.
Without enough documentation or tutorials, it is often challenging even for computer scientists to apply the package to new image data. 
In contrast, UNI-EM~\cite{urakubo2019uni} is scientist-centric, building an integrated system with a graphic interface for both backend deep learning and frontend visualization and manual proofreading.
However, due to the desktop application setup, it is hard to scale UNI-EM to process large-scale data, \eg~petabyte-scale~\cite{shapson2021connectomic,bae2021functional}.
Our PyTC package separates the role of programmers and scientists. For development, PyTC provides design patterns that are scalable and flexible for programmers to build upon. For deployment, PyTC has detailed documentation and step-by-step tutorials for scientists with basic programming experience to use.



\section{Overview}

We give an overview of the framework in this section. Our main goal is to design a modular architecture so that researchers can easily adapt the framework to their data and tasks, and developers can efficiently improve part of the framework (\eg, adding a recent segmentation architecture) without modifying other components for data handling and optimization. Our PyTorch Connectomics (PyTC) package consists of four main modules. The {\em data} module is for reading and pre-processing data at different scales. This module can stream data as tractable volumes for a huge dataset so that the scalability is not restricted by memory. The {\em model} module is to construct 2D and 3D encoder-decoder models with flexibility in customizing the architecture, activations, learning targets, and loss functions. The {\em engine} module supports distributed or non-distributed multi-GPU training and with different combinations of optimizers and learning rate scheduling rules (with optional mixed-precision training), as well as multi-GPU inference.

In addition to the three modules, the {\em configuration} module, which is based on YACS\footnote{\url{https://github.com/rbgirshick/yacs}}, define the configurable system and model hyperparameters like number of GPUs, depth of a segmentation model, and output path. This configuration system ensures reproducibility and version control as it serializes experimental options into YAML files instead of specifying them in the command line. When working on a new dataset, a researcher can quickly update data paths, model architecture, and optimization protocol with the YAML file without touching any source code. More details of those modules are presented in the next section.

\begin{figure}[t]
    \centering
    \includegraphics[width=\columnwidth]{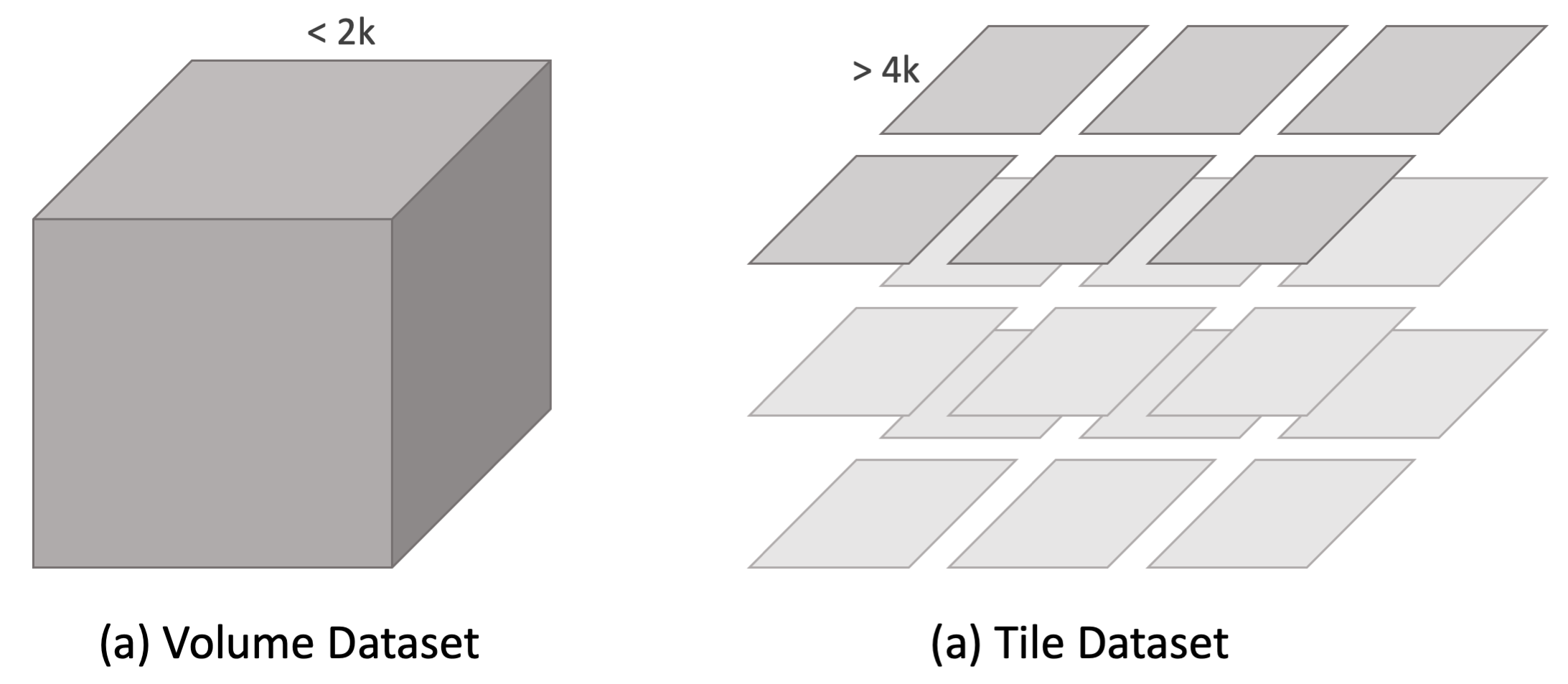}
    \caption{
    Data loading workflow. (a) {\em Volume Dataset} supports random sampling for training and sliding window inference for volumes that can be directly loaded into memory. For large-scale datasets where the volumes are stored as large 2D patches, (b) {\em Tile Dataset} calculates and loads only involved patches to construct tractable volumes. 
    }
    \label{fig:torch_dataset}
\end{figure}
\section{System Design}

The design goals of {\em PyTorch Connectomics} are scalability and flexibility. We focus on data scalability that can handle datasets of different scales, model flexibility for learning multiple targets simultaneously with various loss functions. The system is supported by GPU and CPU parallelism.

\subsection{Data Scalability}\label{sec:data_design}

\bfsection{Data Loading} The growth of connectomics dataset to petabyte scale~\cite{shapson2021connectomic} with various tasks including but not limited to the examples in Figure~\ref{fig:teaser} brings significant challenges in interface standardization. We aim to design a data loading workflow where most components are shared regardless of the data size and computing tasks, so researchers can quickly adapt the system to their data without coding different interfaces. 

Therefore we define the {\em Volume Dataset} class that supports random sampling for training and slide-window inference with overlap for inference (Fig.~\ref{fig:torch_dataset}\red{a}). Optional rejection sampling can be conducted on the fly during training to sample more foreground regions for sparse segmentation tasks like synapses. At inference time, the overlapped sliding-window volumes are combined using a blending function to reduce the border artifacts of CNN models.

Volume Dataset handles volumes at $(5\mu m)^3$ scale (\eg, CREMI~\cite{cremi}) by directly loading them into memory. However, volumes at $(30\mu m)^3$ scale or larger (\eg, MitoEM~\cite{wei2020mitoem}) becomes intractable. Since those datasets are usually saved as large 2D tiles, we designed the {\em Tile Dataset} class (Fig.~\ref{fig:torch_dataset}\red{b}), which reads the file paths and dataset metadata (resolution, patch size, etc.) in initialization. It then crops tractable volumes randomly during training or deterministically during inference to construct a sequence of Volume Dataset instances to conduct the following processing steps. With this design, inference on a large-scale dataset can easily be parallelized by passing different global coordinates to different processes. Even with a limited resource that can handle only one Volume Dataset instance, the framework can work properly (just with more time).

\begin{figure}[t]
    \centering
    \includegraphics[width=\columnwidth]{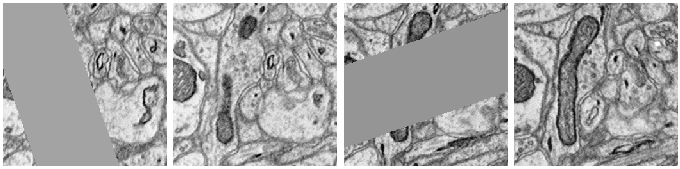}
    \caption{Illustration of missing-part augmentation. As one of over ten training augmentations in PyTC, missing-part augmentation masks out random regions in non-consecutive slices, which force the model to infer the missing structures from the 3D context. Images are consecutive slices.}
    \label{fig:rand_missing}
\end{figure}

\bfsection{Data Augmentation} Since annotating biological and medical datasets requires expertise and time consuming, we developed more than ten different data augmentation techniques in the PyTC framework to improve the robustness of trained segmentation models. The augmentation module consists of non-spatial augmentations that work on images only and spatial augmentations that change images and corresponding segmentation labels simultaneously. Non-spatial augmentations include {\em gray-scale} that randomly adjust contrast and brightness, invert the color space, as well as apply gamma correction, and {\em missing-part} that randomly gray out image regions in non-consecutive slices, which let the model learn to infer the missing structures from the 3D context (Fig.~\ref{fig:rand_missing}). Spatial augmentations include {\em rescale} that randomly scale the image and masks in a pre-specified range, and {\em misalignment} that translates or rotates input volume at some slices to simulate the misalignment problem introduced when stitching 2D patches into 3D volumes during data processing. Details for other segmentation techniques can be found within the package at \texttt{connectomics/data/augmentation}. 

\begin{figure*}[t]
    \centering
    \includegraphics[width=\textwidth]{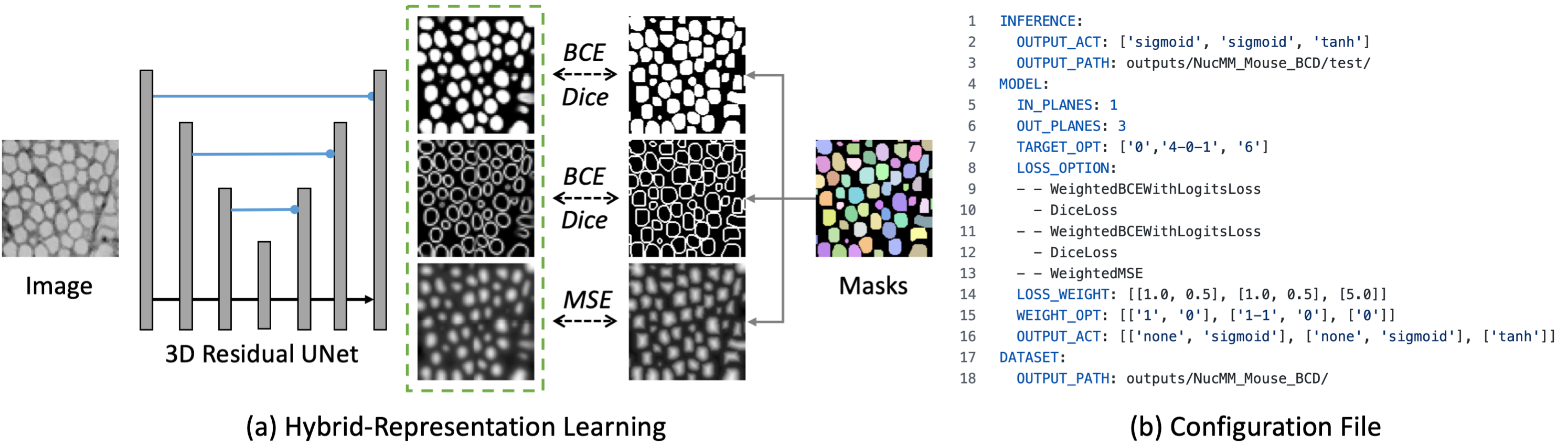}
    \caption{
    Configuration for hybrid-representation learning. (a) The multi-tasking model that learns foreground mask, instance contour, and signed distance transform map simultaneously (with multiple losses for each target) can be easily realized by updating (b) the YAML configuration file. This figure is adapted from Lin \etal~\cite{lin2021nucmm}.
    }
    \label{fig:bcd_yaml}
\end{figure*}

Besides random augmentation at training time, we also support test-time augmentation that runs inference on up to 16 variants (flip along three axes and transpose one $xy$ plane) and average them in a pixel-wise manner after reverting the augmented volumes to the original orientation to generate smooth predictions. Since the augmentation module directly operates on the NumPy array and does not rely on PyTorch, it can be a plug-and-play component for other deep learning and image processing frameworks. 

\subsection{Model Flexibility}

\bfsection{Hybrid-representation learning} We design the framework to handle several semantic and instance segmentation tasks as shown in Figure~\ref{fig:teaser}. In existing work, each task only learns one representation calculated from the ground-truth masks, like the 3D affinity graph for neuron segmentation~\cite{turaga2009maximin,lee2017superhuman,wei2021axonem}, and signed distance transform for synaptic cleft segmentation~\cite{heinrich2018synaptic}. Besides implementing individual learning targets in the package, we design and build a hybrid-representation learning framework where researchers can specify an arbitrary number of targets to learn and an arbitrary number of loss functions for each target by only changing the configuration file without any coding (Fig.~\ref{fig:bcd_yaml}). The weights among different targets and losses are also adjustable based on the downstream application. Our intuition is that simultaneously learning multiple representations using a shared neural network architecture can act as a regularization term to prevent the model from overfitting, and we show in the experiments that this approach performs better for datasets with limited labels.

\bfsection{Active and Semi-supervised Learning}
Data annotation for connectomics requires expertise and is time-consuming. Manual annotation becomes feasible with the skyrocketing data scale~\cite{shapson2021connectomic}, while the model trained on a restricted labeled dataset does not achieve satisfactory performance on vast unlabeled data. Thus besides supervised training, we also incorporate active and semi-supervised learning workflows in this package. Our active learning method is the two-stream active query suggestion algorithm which combines supervised model output and unsupervised feature extractor outputs to suggest the most informative samples in unlabeled data to get new expert annotation~\cite{lin2020two} with a limited budget. For semi-supervised learning, we currently support {\em self-training} that features a two-stage scheme. In the first stage, we train a supervised model on the labeled set. In the second stage, we first run inference on unlabeled data and then combine labeled and pseudo-labeled samples to finetune the model. Despite the simplicity of the approach, recent work has shown significantly improved performance on natural image segmentation benchmarks~\cite{zoph2020rethinking}. We will show in the experiments that self-training also performs well for synaptic cleft segmentation in connectomics. 

\bfsection{Network Architectures}
We designed a highly customizable network architecture so that researchers can freely configure 2D and 3D segmentation models like U-Net~\cite{ronneberger2015u}, DeepLab~\cite{chen2017deeplab} with different numbers of filters, channels, activations, and normalization layers. One main feature of the volumetric image generated by serial-sectioning electron microscopy (EM) is anisotropy: the resolution on the $xy$-plane is 4nm, while 30 to 40nm for the $z$-axis~\cite{kasthuri2015saturated,shapson2021connectomic}. It becomes less reasonable to use symmetric convolutions for all three axes in the model. Therefore the 3D models in our package can also be configured to tackle anisotropic data with mixed 2D and 3D convolutional layers. At an earlier stage of the model, we apply 2D convolutions and only downsample the $xy$-plane with pooling or strided convolution. At a later stage, where the downsampled volume becomes roughly isotropic, we apply symmetric 3D convolutions to make use of context in all three directions.  

\subsection{Training and Inference Parallelism}

\bfsection{Distributed Training} Our framework supports multi-GPU training with {\em DataParallel} where a batch is split and forward separately on different GPUs and aggregated to calculate the loss, as well as {\em DistributedDataParallel} that only communicate gradients between GPUs, powered by PyTorch~\cite{NEURIPS2019_9015}. We make distributed training the default scheme of our package even for a single node as it avoids the overhead caused by the Python Global Interpreter Lock (GIL). For distributed training where multiple processes are running simultaneously with their data loaders, we add one option to control that each process sees a non-overlap part of a dataset, effectively reducing memory usage. This design has another benefit when using together with rejection sampling describe in Sec.~\ref{sec:data_design}. Specifically, some datasets have multiple volumes with different foreground densities (\eg, the density of synapses varies in different brain regions). Thus the loader samples fewer data points from volumes with sparse labels by expectation. However, if we distributed those volumes into different processes in distributed training, the number of sampled data points from multiple volumes becomes the same regardless of the foreground density, avoiding overlooking sparse regions.

\bfsection{Inference Parallelism} Our goal is the framework being scalable in inference with an unlimited number of devices. To achieve this, we design an inference workflow where users can specify a manageable size to chunk a large-scale dataset and update the {\em global} coordinates of the corresponding chunks in the configuration file. All the chunks are then processed in parallel on different nodes as no communication is required between processes. The prediction volumes are stored with an identifier of the global coordinate. By doing this, we can open only relevant predictions for post-processing and analysis, or combine them using a simple post-processing function to reconstruct a volume of the same size as the input image volume\footnote{Example post-processing function for the $(30\ \mu m)^3$ MitoEM dataset is available at \url{https://github.com/zudi-lin/pytorch_connectomics/tree/master/configs/MitoEM}.}.

\begin{table*}[t]
\caption{
Benchmark comparison on the CREMI challenge for synaptic cleft detection. Our semi-supervised learning (SSL) method outperforms existing best result by relatively {\bf 6.1}$\%$ in terms of the overall CREMI score and rank 1st among all challenge submissions. For all metrics in this table, lower is better.
}\label{tab:cremi}

\begin{center}
\small
\setlength{\tabcolsep}{0.5em}
\begin{tabular}{lcccccccccccc}
\toprule
& \multicolumn{3}{c}{Volume A+} & \multicolumn{3}{c}{Volume B+} & \multicolumn{3}{c}{Volume C+} & \multicolumn{3}{c}{{\bf Overall}} \\
Method   & ADGT & ADF & CREMI & ADGT & ADF & CREMI & ADGT & ADF & CREMI & {\bf ADGT} & {\bf ADF} & {\bf CREMI} \\
\midrule
Isensee \etal~\cite{isensee2021nnu} & {\bf 19.02} & 210.58 & 114.80 & 140.41 & 26.73 & 83.57 & {\bf 33.95} & 19.08 & 26.52 & 64.46 & 85.46 & 74.96 \\
Heinrich \etal~\cite{heinrich2018synaptic} & 116.59 & {\bf 28.32} & 72.46 & 176.83 & 30.41 & 103.62 & 35.57 & 17.65 & 26.61 & 109.67 & 25.46 & 67.56 \\
Lin \etal~\cite{lin2020two} & 106.41 & 59.88 & 83.15 & {\bf 134.66} & 22.37 & 78.51 & 51.84 & 8.34 & 30.09 & 96.14 & 33.05 & 64.59 \\
\rowcolor{black!10} {\bf Ours} w/o SSL & 68.62 & 63.63 & 66.13 & 155.40 & {\bf 5.43} & 80.42 & 48.94 & {\bf 5.35} & 27.14 & 90.99 & 24.81 & 57.90 \\
Liu and Ji~\cite{liu2021cleftnet} & 72.60 & 67.11 & 69.85 & 147.50 & 11.52 & 79.52 & 36.69 & 10.97 & {\bf 23.83} & 85.60 & 29.87 & 57.73 \\
\rowcolor{black!10} {\bf Ours} & 68.20 & 60.87 & {\bf 64.53} & 140.62 & 6.41 & {\bf 73.51} & 43.29 & 6.02 & 24.66 & {\bf 84.04} & {\bf 24.43} & {\bf 54.23}\\
\bottomrule
\end{tabular}
\end{center}
\end{table*}
\section{Experiments}

To demonstrate the capability of PyTorch Connectomics, we benchmark the performance of our codebase on two semantic and two instance segmentation datasets covering cellular structures including synapses (Sec.~\ref{sec:cremi}), mitochondria (Sec.~\ref{sec:lucchi} \&~\ref{sec:mitoem}) and neuronal nuclei~(Sec.~\ref{sec:nucmm}).

\subsection{Synaptic Cleft Detection}\label{sec:cremi}

\bfsection{Dataset and Evaluation Metric}
We evaluate PyTC on the CREMI Challenge dataset~\cite{cremi}, which contains three labeled and three unlabeled volumes of the size $1250\times1250\times125$ voxels. The data is collected from adult {\em Drosophila melanogaster} brain tissue with EM at a voxel resolution of $4\times4\times40\ nm$. The results are evaluated by two scores: the average distance of any predicted cleft voxel to its closest ground-truth (ADGT) for penalizing false positives and the average distance of any ground-truth cleft voxel to its closest predicted foreground (ADF) for penalizing false negatives. The final ranking criterion (CREMI score) is the mean of ADGT and ADF over the three test volumes.

\bfsection{Training and Inference}
The model output has a single channel representing the probability synaptic cleft voxels. We first trained a customized 3D U-Net model using an SGD optimizer with linear warmup and cosine annealing in the learning rate scheduler. We use a weighted binary cross-entropy (BCE) loss and applied rejection sampling to reject samples without synapse during training with a probability of $95\%$ to penalize false negatives. The model input size is $257\times257\times17$ in $(x, y, z)$ as CREMI is an anisotropic dataset with higher $x$ and $y$ resolution. The model was optimized for 150K iterations with a batch size of 6 and a base learning rate of 0.02. To further improve the performance, we also use a semi-supervised learning approach called self-training, which generates pseudo-labels on unlabeled images and combines labeled and pseudo-labeled data together in model finetuning. We optimize the model again using the same protocol for 150K iterations.

\begin{figure}[t]
    \centering
    \includegraphics[width=\columnwidth]{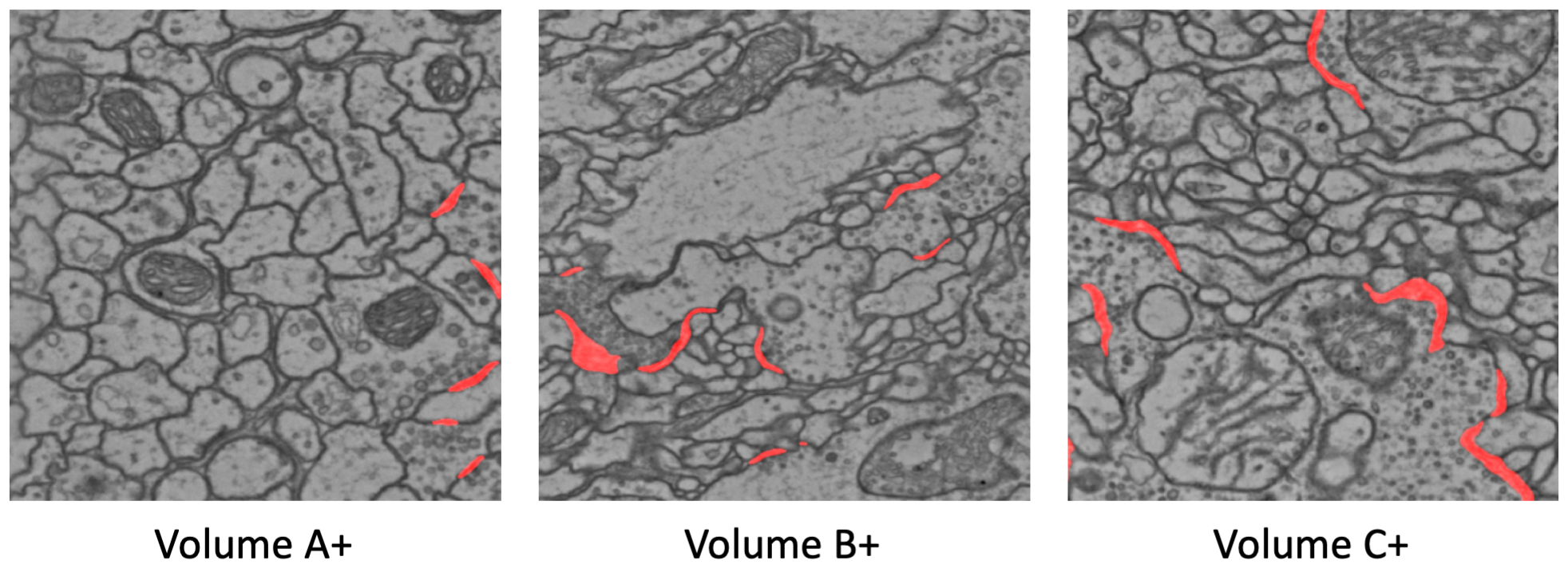}
    \caption{Visualization of synaptic cleft prediction (red masks) on the CREMI challenge test volumes, which are EM images of the adult fly brain.}
    \label{fig:cremi}
\end{figure}

\bfsection{Results}
Table~\ref{tab:cremi} shows comparison with results reported in existing publications, including nnUNet~\cite{isensee2021nnu} that features a self-adapting framework based on the original U-Net~\cite{ronneberger2015u}, Heinrich \etal~\cite{heinrich2018synaptic} that learns a signed distance transform, Lin \etal~\cite{lin2020two} that employs an asymmetric U-Net architecture, and CleftNet~\cite{liu2021cleftnet} that augments both features and labels for better learning. Our binary segmentation model without semi-supervised learning achieves an overall CREMI score of 57.90, which is comparable with the performance of the existing state-of-the-art approach at 57.73~\cite{liu2021cleftnet}. Our final model achieves the best overall performance in terms of ADGT, ADF, and CREMI scores. For the overall CREMI score, which is the ranking criterion of the challenge, our result significantly improves previous state-of-the-art by relatively $6.1\%$. We also visualize the prediction on the test volumes in Fig.~\ref{fig:cremi}. Even including all results on the CREMI leaderboard, our method still ranks 1st among the challenge submissions (by Dec 09, 2021).

\subsection{Mitochondria Semantic Segmentation}\label{sec:lucchi}
\begin{figure}[t]
    \centering
    \includegraphics[width=\columnwidth]{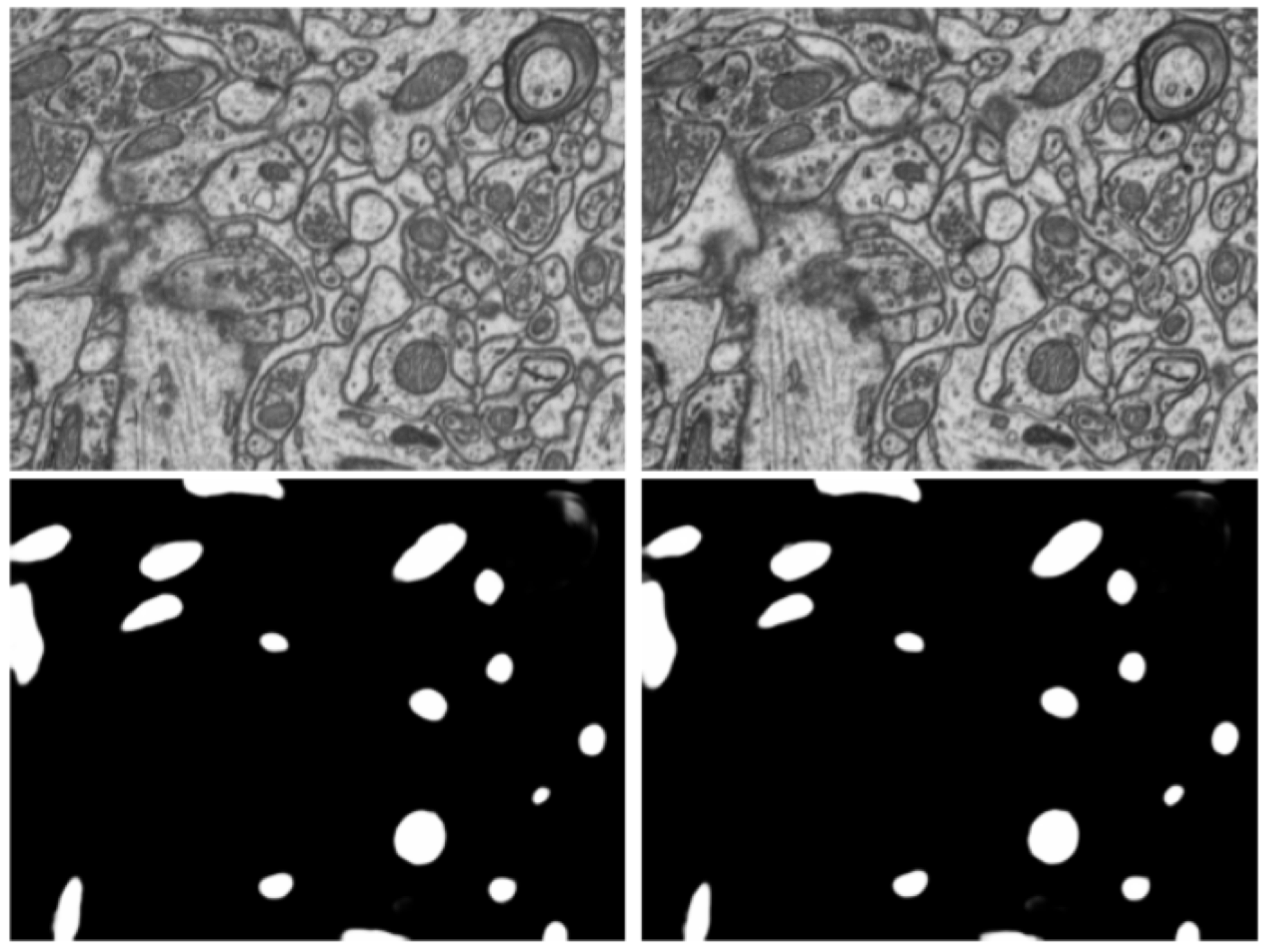}
    \caption{Visualization of model prediction (foreground probability) on the mitochondria segmentation dataset released by Lucchi \etal~\cite{lucchi2012structured}, without any post-processing.}
    \label{fig:lucchi}
\end{figure}
\bfsection{Dataset and Evaluation Metric}
For mitochondria semantic segmentation, we compare the model implemented with our PyTC framework with previous approaches on the EPFL Hippocampus dataset (also widely known as the Lucchi dataset)~\cite{lucchi2012structured}, which contains one training and one test volumes of size $1024\times768\times165$ voxels. Following previous work~\cite{casser2020fast}, we evaluate the prediction using both foreground IoU and overall IoU. Casser \etal~\cite{casser2020fast} provides the annotation from a neuroscience expert, whose performance is compared against the official annotation to define the {\em human annotation} performance on this task.
\begin{table}[t]
\caption{Benchmark comparison on the mitochondria semantic segmentation dataset released by Lucchi \etal~\cite{lucchi2012structured}. The results are evaluated by both foreground IoU and overall IoU (higher is better). Our model outperforms human annotation without any complex post-processing.}
\begin{center}\label{tab:lucchi}
\begin{tabular}{lcc}
\toprule
Method & FG-IoU & IoU \\ 
\midrule
Oztel \etal~\cite{oztel2017mitochondria} & 0.907 & -  \\
Xiao \etal~\cite{xiao2018automatic} & 0.900 & - \\ 
Lucchi \etal~\cite{lucchi2015learning} & 0.895 & 0.948  \\
\rowcolor{black!10} Ours & 0.892 & 0.943 \\
Casser \etal~\cite{casser2020fast} & 0.890 & 0.942 \\
Cheng and Varshney~\cite{cheng2017volume} & 0.889 & 0.942 \\
{\bf Human Annotation*}~\cite{casser2020fast}  & 0.884 & 0.938 \\
Cetina \etal~\cite{cetina2018multi} & 0.760 & - \\
M{\'a}rquez-Neila \etal~\cite{marquez2014non} & 0.762 & -   \\
Lucchi \etal~\cite{lucchi2014exploiting} & 0.741 & -  \\
\bottomrule
\end{tabular}
\end{center}
\end{table}

\bfsection{Training and Inference}
Similar to the CREMI experiments, the model output is the probability map of mitochondria. Since this dataset is isotropic (each voxel is a cube), we use an input size of $112\times 112\times 112$, and we only use 3D convolutional filters in our custom 3D U-Net architecture instead of a combination of 2D and 3D kernels as for anisotropic data. For training augmentation, we enable transpose between every pair of three axes as the input is cubic. We use both weighted BCE and Dice losses with a ratio of 1:1 and train the model for 100K iterations with a batch size of 8. For post-processing, we applied median filtering with a kernel size of $7\times7\times7$.

\bfsection{Results} Fig.~\ref{fig:lucchi} visualizes the raw model prediction without thresholding or median filtering on the EPFL Hippocampus test volume. In the quantitative comparison with previous state-of-the-art works on this dataset (Table~\ref{tab:lucchi}), we show that our model surpasses human annotation performance provided by Casser \etal~\cite{casser2020fast}. Besides, different from the best performing method~\cite{oztel2017mitochondria} which has a complex post-processing step including watershed-based boundary refinement, we only apply a simple median filtering with a cubic kernel after model inference. 

\subsection{Mitochondria Instance Segmentation}\label{sec:mitoem}
\begin{figure}[t]
    \centering
    \includegraphics[width=\columnwidth]{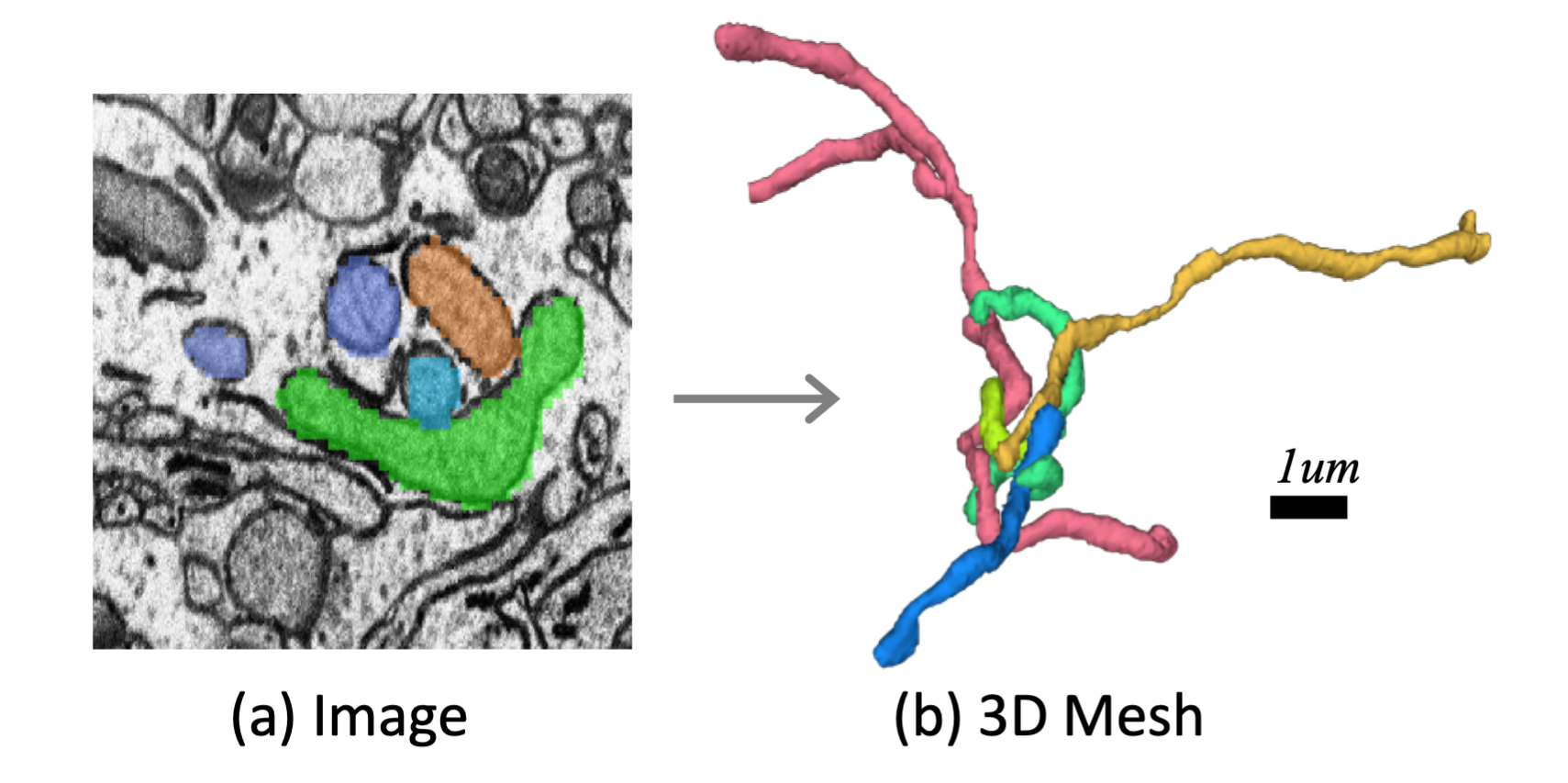}
    \caption{Challenging cases in MitoEM. We show the image and 3D meshes of a dense tangle pattern of touching mitochondria. This figure is adapted from Wei \etal~\cite{wei2020mitoem}.}
    \label{fig:mitoem}
\end{figure}

\bfsection{Dataset and Evaluation Metric}
Besides semantic segmentation tasks, including synaptic cleft and mitochondria segmentation, we also work on the more challenging instance segmentation problem, which assigns each object a unique index. Specifically, we use the large-scale MitoEM dataset~\cite{wei2020mitoem} for segmenting mitochondria instances, which is over $3,600\times$ larger than the Lucchi dataset~\cite{lucchi2012structured} and has two volumes covering human (MitoEM-H) and rat (MitoEM-R) brain tissues, respectively. A difficult case for segmentation algorithms is demonstrated in Fig.~\ref{fig:lucchi}. Both volumes are partitioned into consecutive train, val and test splits with 40\%, 10\% and 50\% of the data. For evaluation, we use the 3D average precision (AP) metric with a mask IoU threshold of 0.75, denoted as AP-75. The test sets are automatically evaluated on the challenge website\footnote{IEEE-ISBI 2021 MitoEM challenge: \url{https://mitoem.grand-challenge.org/}}.

\bfsection{Training and Inference}
We use the {\em Tile Dataset} to process MitoEM as the volumes are too large to be directly loaded into memory. Different from the binary semantic segmentation models that have only one output channel, we build a U3D-BC architecture~\cite{wei2020mitoem} that predicts the foreground mask and instance contour map at the same time. The instance contour map is informative in separating closely touching instances. We train the models for 150K iterations from scratch with a batch size of 8 and an input size of $257\times257\times17$ as the data is also anisotropic. We train two models for two volumes separately in this comparison. The watershed segmentation algorithm is applied after merging overlapping chunks back to a single volume.

\begin{table}[t]
\caption{Benchmark comparison on the MitoEM challenge dataset~\cite{wei2020mitoem} for mitochondria instance segmentation. The results are AP-75 scores evaluated on two volumes. Our model ranks 3nd among challenge participants.}
\begin{center}\label{tab:mitoem}
\begin{tabular}{lccc}
\toprule
Method & MitoEM-R & MitoEM-H & Overall \\ 
\midrule
M. Li \etal~\cite{li2021advanced} & 0.851 & 0.829 & 0.840 \\ 
\rowcolor{black!10} U3D-BC (Ours) & 0.816 & 0.804 & 0.810 \\
Z. Li \etal & 0.815 & 0.783 & 0.799 \\
C. Pape & 0.770 & 0.770 & 0.770 \\
R. Conrad & 0.662 & 0.679 & 0.671 \\
\bottomrule
\end{tabular}
\end{center}
\end{table}

\bfsection{Results} On the MitoEM challenge dataset, our model achieves an AP-75 score of 0.816 for the rat volume and 0.804 for the human volume, which is on average 0.810 and ranks 2nd among the submissions from challenge participants. We noticed that the segmentation performance for small instances ($n<$5K voxels) are relatively low (0.311 for rat and 0.426 for human, respectively). We argue this is because small objects are much easier to be ignored than medium (5K$<n<30$K voxels) and large ($n>$30K voxels) ones with a similar thickness of predicted instance boundaries. Therefore for future work, we will work on improving the segmentation for small instances without sacrificing accuracy for medium and large instances. 

\subsection{Neuronal Nuclei Instance Segmentation}\label{sec:nucmm}

\bfsection{Dataset and Evaluation Metric}
We use the NucMM dataset~\cite{lin2021nucmm} that has one EM volume of zebrafish brain (Fig.~\ref{fig:teaser}\red{d}) and one micro-CT volume of mouse visual cortex. The numbers of neuronal nuclei are 170K and 7K for two volumes, respectively. Unlike those datasets described above with nanometer resolution, the two volumes have almost isotropic voxels of $(0.5\ \mu m)^3$ and $(0.7\ \mu m)^3$, respectively. Each volume is split into $5\%$ training set, $5\%$ validation set, and $90\%$ test set. For evaluation, we use the 3D average precision (AP) metric similar to the MitoEM data for mitochondria instance segmentation~\cite{wei2020mitoem}. However, instead of using one single IoU threshold of 0.75, we evaluated at both AP-50 and AP-75 as the objects are relatively small (not physically but in terms of the number of voxels).

\bfsection{Training and Inference}
We construct two hybrid-representation learning models, including a U3D-BC that learns foreground mask and instance contour at the same time (similar to mitochondria instance segmentation) and a U3D-BCD model that additionally learns a signed distance map (Fig.~\ref{fig:bcd_yaml}). Specifically, the distance map is
\begin{equation}
  f(x_i)=\begin{cases}
    +\text{dist}(x_i, B) / \alpha, & \text{if $x\in F$}.\\
    -\text{dist}(x_i, F) / \beta, & \text{if $x\in B$}.
  \end{cases}
\end{equation}
where $F$ and $B$ denote the foreground and background regions, while $\alpha$ and $\beta$ control the scale of the distance.
For U3D-BC, we use the default 1.0 weight ratio between the foreground and contour map losses. For U3D-BCD, we set $\alpha$ and $\beta$ of the signed distance map to 8 and 50, respectively, without tweaking. The segmentation models are trained separately on both volumes because the image appearances and nuclei distributions are quite distinct. For the U3D-BCD model, all three model predictions are used in the watershed segmentation step to yield the instance masks.

\bfsection{Results}
We show the AP scores (AP-50, AP-75, and their average) on the 90$\%$ test sets for both volumes after hyperparameter tunning on the validation sets (Table~\ref{tab:nucmm}). The overall performance is the mean over two NucMM volumes. The results show that both our U3D-BC and U3D-BCD models compare favorably against existing nuclei segmentation methods, including Cellpose~\cite{stringer2021cellpose} and Stardist~\cite{weigert2020star}. Besides, the U3D-BCD model that learns an additional distance map outperforms the U3D-BC model by relatively $22\%$ in overall performance. We argue this is because the signed distance map preserves more information of the instance structure and has more supervision on the background regions as the other two representations treat background pixels close or distant from the foreground equally.

\begin{table}[t]
\caption{\label{tab:nucmm}
Benchmark comparison on the NucMM dataset. We compare state-of-the-art methods on the NucMM dataset using the average precision (AP) metric. {\bf Bold} and \underline{underlined} numbers denote the 1st and 2nd scores, respectively. Both U3D-BC~\cite{wei2020mitoem} and U3D-BCD~\cite{lin2021nucmm} models are implemented with our PyTC codebase.
}

\begin{center}
\small
\setlength{\tabcolsep}{0.5em}
\begin{tabular}{lccccc}
\toprule
\multirow{2}[2]{*}{Method}
& \multicolumn{2}{c}{NucMM-Z}
& \multicolumn{2}{c}{NucMM-M}
& \multirow{2}[2]{*}{Overall}
\\\cmidrule(lr){2-3} \cmidrule(lr){4-5}
& AP-50 
& AP-75 
& AP-50 
& AP-75\\

\midrule

Cellpose~\cite{stringer2021cellpose}
& 0.796 & 0.342 & 0.463 & 0.002 & 0.401\\
StarDist~\cite{weigert2020star}
& \underline{0.912} & 0.328 & 0.306 & 0.004 & 0.388\\
\rowcolor{black!10} U3D-BC~\cite{wei2020mitoem}
& 0.782 & \underline{0.556} & {\bf 0.645} & \underline{0.210} & \underline{0.549}\\

\rowcolor{black!10} U3D-BCD~\cite{lin2021nucmm}
& {\bf 0.978} & {\bf 0.809} & \underline{0.638} & {\bf 0.250} &
{\bf 0.669}\\
\bottomrule
\end{tabular}
\end{center}
\end{table}
\section{Discussion and Future Work} In this paper, we present the {\em PyTorch Connectomics} deep learning toolbox that is flexible in constructing appropriate models for diverse tasks and scalable in tackling volumetric microscopy datasets at different sizes. Since connectomics (and the more general microscopy image analysis) is a fast-growing field involving multiple imaging modalities and generates image data at petabyte scales, there are many opportunities and challenges calling for effort from the computer vision and machine learning aspects. For the next step, we identify three directions that will improve the capability of our framework in assisting researchers in biological and medical analysis.

\bfsection{Unsupervised Visual Representation Learning} Recent work on unsupervised representation learning from raw images without any labels~\cite{he2020momentum,chen2020simple} has achieved impressive performance in learning meaningful representations, which tackles diverse downstream tasks by finetuning on a small amount of labeled data. The most widely used idea is contrastive learning, which makes two augmentations of the same image close and different images apart in the feature space under some similarity measure (\eg, cosine similarity). Considering the diverse modalities and large scale of connectomics data and the expensiveness in obtaining expert annotations, building generic architectures by unsupervised learning on raw images can be a promising way to improve the finetuning performance for specific supervised learning tasks. It can also be applied together with active learning and semi-supervised learning to utilize vast unlabeled images effectively under an annotation budget.

\bfsection{Improving Training and Inference Efficiency} 
With the use of multi-beam electron microscopes that image a hexagonal area of about $10,000 \mu m^2$ simultaneously at nanometer resolution~\cite{shapson2021connectomic}, the training and inference of deep learning models have become a bottleneck in the connectomics workflow. For inference, besides parallel processing on multiple nodes with multiple GPUs, we plan to incorporate recent techniques including network distillation~\cite{hinton2015distilling}, pruning~\cite{han2015learning} and quantization~\cite{courbariaux2015binaryconnect} to significantly improve the inference efficiency without a performance drop. For training, we will implement the multigrid method designed for efficiently optimizing video models~\cite{wu2020multigrid}. The unsupervised visual representation learning approach described above can also improve training efficiency as fewer iterations are needed to achieve similar or higher performance by starting from such pretrained models.

\bfsection{Semi-automatic Proofreading} Proofreading is a crucial step in data-driven biomedical research because current automatic algorithms do not generate satisfying segmentation results for downstream morphology and network analysis. Without easy-to-use software, both the proofreading and the training of annotators can be time-consuming. Our idea for achieving high proofreading efficiency is to develop robust learning algorithms that can suggest errors in automatic predictions and provide choices of high-quality corrections to effectively decrease the navigation time and the number of required interactions, respectively. For suggesting errors, we will use semi-supervised learning to capture mispredictions in unlabeled data by incorporating them in model optimization. For error correction, we plan to build a shape-representation model that can suggest the fixation of errors based on shape priors learned from annotated masks.

\subsection*{Acknowledgement} Our PyTorch Connectomics is built upon numerous previous projects in the connectomics and computer vision community. Especially, we would like to thank the contributors of the following GitHub repositories: PyGreentea\footnote{\url{https://github.com/naibaf7/PyGreentea}}, DataProvider\footnote{\url{https://github.com/torms3/DataProvider}} and Detectron2\footnote{\url{https://github.com/facebookresearch/detectron2}}. We thank Jason Adhinarta, Siddharth Awasthi, Atmadeep Banerjee, Mourad Belhamissi, Aarush Gupta, Krishna Swaroop K, Emil Krauch, Aayush Kumar, Yuhao Lu, Atharva Peshkar, Pragya Singh, Qijia Shen, Karan Uppal, Zijie Zhao, Chenfan Zhuang, and Silin Zou (sorted by last name) for contributing to this package. We thank our close collaborators Ignacio Arganda-Carreras and Xueying (Snow) Wang for the exciting explorations together. We thank the co-authors of our connectomics-related publications for the effort to make the field more accessible to computer science researchers. Finally, we gratefully acknowledge the support from NSF awards IIS-1835231 and IIS-2124179.

{\small
\bibliographystyle{ieee_fullname}
\bibliography{main}
}

\end{document}